\documentclass[conference]{IEEEtran}
\IEEEoverridecommandlockouts
\usepackage{amsmath,amssymb}
\usepackage{graphicx}
\usepackage{cite}
\usepackage{hyperref}
\usepackage[utf8]{inputenc}
\hypersetup{draft} 

\usepackage{algorithm}
\usepackage{algorithmicx}
\usepackage{algpseudocode}
\usepackage{tikz}
\usetikzlibrary{arrows.meta, positioning}
\usepackage{bm}
\usepackage{comment}
\usepackage{color}

\newcommand{\ignore}[1]{}

\title{MAPPO for Edge Server Monitoring \\

\thanks{This work has been supported in part by the Army Research Laboratory and was accomplished under Cooperative Agreement Number W911NF-24-2-0205, and by the U.S. National Science Foundation under the grant CNS-2239677. The views and conclusions contained in this document are those of the authors and should not be interpreted as representing the official policies, either expressed or implied, of the Army Research Laboratory or the U.S. Government. The U.S. Government is authorized to reproduce and distribute reprints for Government purposes notwithstanding any copyright notation herein.}
}

\author{
\IEEEauthorblockN{Samuel Chamoun\dag, Christian McDowell\dag, Robin Buchanan\dag, Kevin Chan\ddag, Eric Graves\ddag,  and Yin Sun\dag}  
\IEEEauthorblockA{\dag{Department of Electrical and Computer Engineering, Auburn University, Auburn, AL, USA} \\
\ddag
{DEVCOM Army Research Laboratory, Adelphi, MD, USA}\\ 
}
}
\begin{document}

\maketitle

\begin{abstract}
    In this paper, we consider a goal-oriented communication problem for edge server monitoring, where jobs arrive intermittently at multiple dispatchers and must be assigned to shared edge servers with finite queues and time-varying availability. Accurate knowledge of server status is critical for sustaining high throughput, yet remains challenging under dynamic workloads and partial observability. To address this challenge, each dispatcher maintains server knowledge through two complementary mechanisms: (i) active status queries that provide instantaneous updates at a communication cost, and (ii) job execution feedback that reveals server conditions upon successful or failed job completion. We formulate a cooperative multi-agent distributed decision-making problem in which dispatchers jointly optimize query scheduling to balance throughput against communication overhead. To solve this problem, we propose a Multi-Agent Proximal Policy Optimization (MAPPO)-based algorithm that leverages centralized training with decentralized execution (CTDE) to learn distributed query-and-dispatch policies under partial and stale observations. Experiments show that MAPPO achieves superior throughput-cost tradeoffs and significantly outperforms baseline strategies across varying query costs, job arrival rates, and dispatchers.
\end{abstract}

\section{Introduction}
In edge computing systems deployed in contested or resource-constrained environments, efficient monitoring of edge servers and timely job dispatching are critical for sustaining mission-critical real-time edge services such as Intelligence, Surveillance, and Reconnaissance (ISR), enhanced situational awareness in dynamic operational theaters, and coordinated control of autonomous systems in multi-agent missions. For instance, a tactical surveillance system processing live video streams from ground and aerial sensors may generate bursts of compute-intensive jobs that must be routed to a limited set of edge AI servers. However, server availability is often uncertain and highly dynamic, influenced by competing computational workloads, intermittent background tasks, and time-varying job arrivals. Conventional systems typically rely on periodic polling to retrieve status updates from edge servers and support job dispatching decisions, but this incurs significant query overhead and scales poorly when each dispatcher needs to retrieve the statuses from all servers. 

In this paper, we consider a goal-oriented communication problem for edge server monitoring, where jobs arrive intermittently at multiple dispatchers and must be assigned to shared edge servers with finite queues and time-varying availability. 
Each dispatcher maintains server knowledge through two feedback mechanisms: (i) active queries, which provide instantaneous availability and queue length at a communication cost, and (ii) job execution feedback, which passively reveals the same information upon successful or failed job completion. These mechanisms differ in cost and timeliness: queries yield controllable updates, whereas job execution feedback is  opportunistic. This creates a tradeoff between sustaining throughput and limiting query overhead, especially in contested environments with limited bandwidth and energy resources. The technical contributions of this paper are summarized as follows:
\begin{itemize}
    \item We optimize server query decisions to balance the trade-off between job throughput and query cost. This trade-off depends on the dispatcher’s knowledge of server availability and queue length. However, because query opportunities are constrained by limited channel resources, this knowledge can become outdated, leading to aged information. To capture this effect, we formulate the problem within an Age of Information (AoI)-aware framework, where the freshness of server state information directly influences decision-making and overall system performance.
    \item Query scheduling is formulated as a Cooperative Multi-Agent Partially Observable Markov Decision Process (MA-POMDP) and solved using a MAPPO-based algorithm. The algorithm leverages centralized training with decentralized execution (CTDE), enabling dispatchers to learn distributed query-and-dispatch policies under partial and stale observations.
    \item Numerical evaluations demonstrate that the proposed MAPPO-based algorithm consistently outperforms conventional baselines in job throughput and query efficiency across varying query costs, job arrival rates, and numbers of dispatchers, underscoring the advantage of jointly optimizing querying and dispatching decisions.
\end{itemize}

\section{Related Work}

Prior works have explored how to monitor dynamic computing environments under communication and resource constraints. In tactical networks, adaptive monitoring strategies were proposed to support analytics placement across edge nodes~\cite{corcoran2022adaptive}, while distributed update policies were studied for tracking remote sources with limited bandwidth~\cite{graves2024optimal}. Recent efforts extended these ideas to edge AI, where monitoring compute resources is critical for distributed learning~\cite{chamoun2025milcom}. Complementary approaches addressed service placement, request scheduling, and adaptive migration using MDP-based formulations~\cite{wang2018edge,wang2019adaptive}, mobility-aware policies~\cite{wang2019dynamic}, and job scheduling in Markovian systems~\cite{banerjee2025tracking}. While these studies highlight the importance of timely state tracking, most assume centralized coordination or static system knowledge, leaving open challenges in decentralized multi-dispatcher systems where monitoring and job assignment are jointly coupled.

To overcome the limitations of delayed or incomplete server knowledge in decentralized systems, the Age of Information (AoI) has been established as a key metric for quantifying the freshness of state updates~\cite{kaul2012real}. Early work optimized AoI through queueing and scheduling policies~\cite{kaul2012real, sun2016update, sun2017update, kadota2018optimizing}, while subsequent studies introduced information-theoretic metrics such as mutual information and feature–time correlation to better capture update utility~\cite{sun2018information, sun2019sampling, Shisher2024Timely, ChakrabortySubmitted2025}. Building on these foundations, extensions have considered the role of AoI in communication systems, such as leveraging the age of channel state information to improve throughput performance~\cite{chakraborty2025send}. AoI-aware estimation studies further established how stale updates degrade remote tracking accuracy under both Markovian~\cite{sun2017remote} and autoregressive models~\cite{shisher2021age, shisher2024monotonicity}. Empirical results confirmed that outdated features impair real-time inference and decision-making~\cite{tsai2021unifying, shisher2023learning, chakraborty2025timely}.

The challenge of coordinating job dispatching under stale and partial observations naturally motivates multi-agent reinforcement learning (MARL). Proximal Policy Optimization (PPO)~\cite{schulman2017proximal} has emerged as a leading reinforcement learning method for its empirical stability and ease of implementation, with follow-up work analyzing key implementation factors~\cite{engstrom2019implementation}. To support cooperative settings, Multi-Agent PPO (MAPPO) extends PPO to leverage centralized training with decentralized execution, showing strong performance across multi-agent benchmarks~\cite{yu2022surprising}. MAPPO has since been applied to domains including UAV coordination and disaster recovery~\cite{kang2023cooperative, guan2023mappo}, with surveys highlighting its growing impact on robotics and autonomy~\cite{orr2023multi, ebochi2025survey}. While these studies demonstrate the effectiveness of MAPPO across diverse domains, none have address decentralized job dispatching under AoI-constrained observations, where explicit server queries and implicit execution feedback must be jointly optimized in a partially observable environment.

\section{System Model}

\begin{figure}[t]
    \centering
    \includegraphics[width=0.8\linewidth]{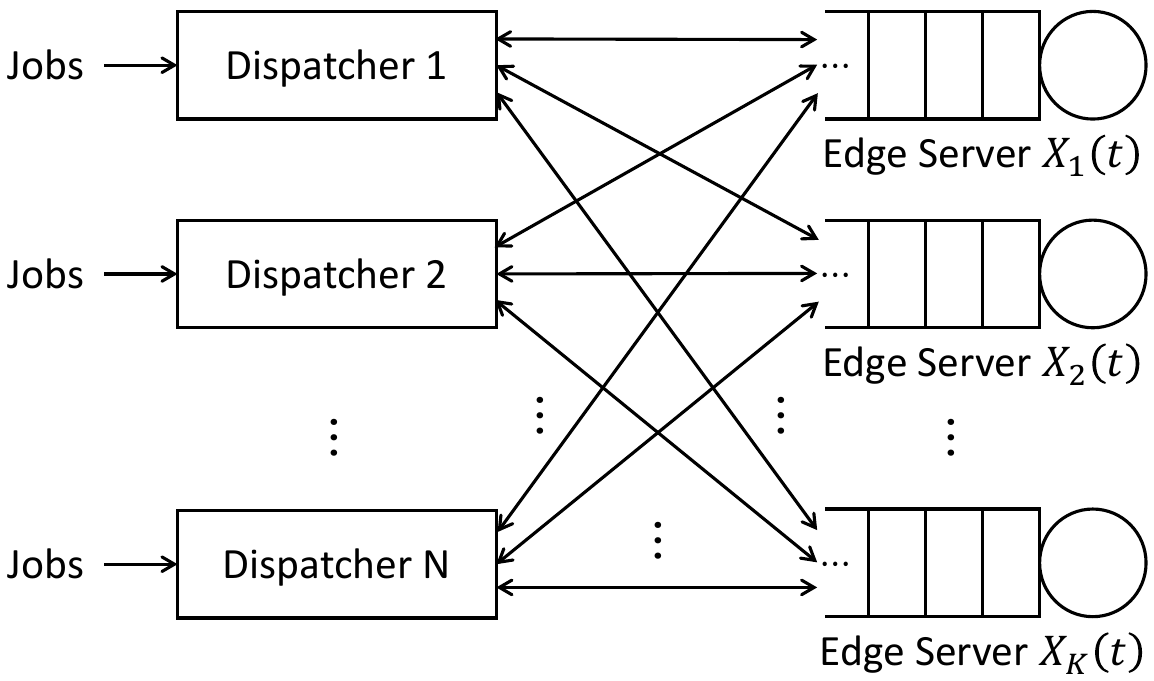}
    \caption{System model: Jobs arrive intermittently at each of the $N$ dispatchers and are immediately assigned to one of the $K$ shared edge servers. Each server maintains a FIFO queue and may become temporarily unavailable due to background tasks. To estimate server availability, each dispatcher relies on two types of feedback: (i) active queries sent to servers, and (ii) passive feedback received from past job executions.}
    \label{fig:system_model}
\end{figure}

\subsection{Job Service and Queueing Model}
Consider a time-slotted edge computing system, consisting of \(N\) job dispatchers, indexed by \(n \in \{1,\dots,N\}\), and \(K\) edge servers, indexed by \(k \in \{1,\dots,K\}\), as illustrated in Figure~\ref{fig:system_model}, where the $K$ servers are shared among all $N$ dispatchers. Each dispatcher can assign incoming jobs to any server, and each server can complete at most one job per time slot. 

The job arrivals of each dispatcher $n$ are modeled as a Bernoulli process. Let \(z_n(t) \in \{0,1\}\) be a Bernoulli random variable such that \(z_n(t) = 1\) if a job arrives at dispatcher \(n\) at time \(t\), and \(z_n(t) = 0\) otherwise. The random variables \(z_n(t)\) are independent across dispatchers and \emph{i.i.d.} across time slots, with mean  \(\mathbb E[z_n(t)] = \lambda_n\). If a job arrives at the beginning of time slot $t$, i.e., $z_n(t) = 1$, dispatcher~$n$ immediately assigns it to one of the $K$ servers; if \(z_n(t) = 0\), no dispatch occurs. 

Due to other background computation tasks on the edge servers, the availability of each server for processing the incoming jobs varies over time. 
Let $X_{k}(t) \in \{0,1\}$ represent the availability status of the $k$-th server, where
\begin{align}
   \!\!\!\!\!X_{k}(t)=
   \begin{cases}
       0, & \text{if server } k \text{ is available at time slot } t,\\
       1, & \text{if server } k \text{ is unavailable at time slot } t.
   \end{cases}
   \label{eq:server_status}
\end{align}
We assume that \(X_{k}(t)\) evolves as an ergodic binary Markov chain with transition matrix 
\begin{IEEEeqnarray}{rCl}
   \begin{bmatrix}
       \phi_{k} & 1 - \phi_{k} \\
       1 - \psi_{k} & \psi_{k}
   \end{bmatrix} = \bm P_{k},
   \label{eq:transition_matrix}
\end{IEEEeqnarray}
where \(\phi_{k} \in (0,1)\) represents the probability that server \(k\), if available for processing an incoming job at time slot \(t\), remains available at time slot \(t+1\); and \( \psi_{k}\in(0,1)\)  is the probability that server \(k\), if unavailable at time slot \(t\), remains unavailable at time slot \(t+1\). Let $\boldsymbol{\pi}_{k} = \begin{bmatrix} \pi_{k}^{(0)} & \pi_{k}^{(1)} \end{bmatrix}$ denote the stationary distribution of the Markov chain $X_{k}(t)$, which can be obtained by solving the following equations:
\begin{IEEEeqnarray}{rCl}
\boldsymbol{\pi}_{k} \bm P_{k} = \boldsymbol{\pi}_{k}, \quad \pi_{k}^{(0)} + \pi_{k}^{(1)} = 1.
\label{eq:stationary_distribution}
\end{IEEEeqnarray}
The system starts to operate at time slot $t = 0$, with $X_{k}(0)$ assumed to follow the stationary distribution $\boldsymbol{\pi}_{k}$.

Each server $k$ maintains a First-In, First-Out (FIFO) queue with a finite buffer size \(Q_k\) for storing incoming jobs from the dispatchers. Let \(q_k(t) \in \{0,1,\dots,Q_k\}\) denote the queue length at the beginning of slot \(t\), defined as the number of jobs in the buffer including any job currently in service. 
When dispatcher \(n\) sends a job to server \(k\), it is appended to the queue; if the queuing buffer is full, the oldest job in the queue is dropped to maintain the finite buffer size \(Q_k\). %
If the queue is nonempty and the server is available (\(X_k(t) = 0\)), one job is processed in slot \(t\). The queue length evolves as
\begin{IEEEeqnarray}{rCl}
q_k(t+1) &=& \min\Big\{ \big[ q_k(t) + \sum_{n=1}^N D_{n,k}(t) \nonumber \\
& &\qquad - S_k(t) \big]^+,\, Q_k \Big\}.
\label{eq:queue_update}
\end{IEEEeqnarray}
where $[x]^+ = \max\{x,0\}$, \(D_{n,k}(t) \in \{0,1\}\) equals 1 if dispatcher \(n\) sends a job to server \(k\) in slot \(t\), and 0 otherwise;  
and \(S_k(t) = 1- X_k(t)\), that is,  $S_k(t) = 1$ if server \(k\) is available (\(X_k(t) = 0\)), and 0 otherwise.

\subsection{Server Status Query and Job Feedback Model}
Dispatchers update their knowledge of server availability using both (i) active server status queries and (ii) job processing acknowledgments from servers. On the one hand, if dispatcher $n$ sends a query to server $k$, it returns the current availability state \(X_k(t)\) and queue length \(q_k(t)\). Let \(c_{n,k}(t) \in \{0,1\}\) denote whether dispatcher \(n\) queries server \(k\) at time \(t\), defined as
\begin{IEEEeqnarray}{rCl}
c_{n,k}(t) & = &
\begin{cases}
1, &
\begin{array}[t]{@{}l@{}}
\text{if dispatcher $n$ queries server $k$} \\
\text{at time $t$,}
\end{array} \\
0, & \text{otherwise.}
\end{cases}
\label{eq:query_def}
\end{IEEEeqnarray}
Each dispatcher $n$ can send multiple queries to different servers in the same time slot $t$. 
Each query incurs a cost $\beta \geq 0$, as explained in Section \ref{sec:reward}, and is assumed to return the server’s current state within the same time slot.
On the other hand, a job processing acknowledgment from server $k$ to dispatcher $n$ conveys the same information \(X_k(t)\) and \(q_k(t)\), but only after the corresponding job has been completed or dropped. Specifically, server $k$ sends an ACK when it completes a job from dispatcher $n$, and an NAK if the job is dropped due to buffer overflow; both events may occur several time slots after the job was dispatched.
Let \(\alpha_{n,k}(t)\) denote the acknowledgment feedback from server \(k\) to dispatcher \(n\), defined as
\begin{IEEEeqnarray}{rCl}
\!\alpha_{n,k}(t) & = &
\begin{cases}
(j,\, X_k(t),\, \sigma_{n,k}(t)), &
\!\begin{array}[t]{@{}l@{}}
\!\text{if dispatcher $n$ receives} \\
\!\text{an ACK/NAK for} \\
\!\text{job $j$ from server $k$,}
\end{array} \\
0, & \!\!\text{otherwise.}
\end{cases} 
\label{eq:ack_def}
\end{IEEEeqnarray}
where \(\sigma_{n,k}(t) \in \{0,1\}\) is the job completion flag, with \(\sigma_{n,k}(t) = 0\) indicating the job was rejected (NAK) due to queue overflow and \(\sigma_{n,k}(t) = 1\) indicating the job was accepted (ACK).

If dispatcher \(n\) receives a status update from server \(k\) by the end of a slot, either through a query \((c_{n,k}=1)\) or an acknowledgment \((\alpha_{n,k} \neq 0)\), its AoI is reset to one at time slot $t+1$; otherwise, the AoI increases by one. Hence, the AoI evolves according to
\begin{IEEEeqnarray}{rCl}
\Delta_{n,k}(t+1)\! =\!
\begin{cases}
1, & \text{if } \alpha_{n,k}(t) \!\neq\! 0 \text{ or } c_{n,k}(t)\! = \!1, \\
\Delta_{n,k}(t) + 1,\!\!\!\! & \text{otherwise},
\end{cases}~~~
\label{eq:aoi_update}
\end{IEEEeqnarray}

\section{Cooperative MA-POMDP Problem Formulation}
\label{sec:pomdp}
In this section, we formulate a joint optimization problem for job dispatching and server querying. Each dispatcher makes decisions separately, based on its knowledge on server availability obtained through server querying and job processing acknowledgment. This problem is formulated as a cooperative Multi-Agent Partially Observable Markov Decision Process (MA-POMDP), specified by a 7-tuple 
$\mathcal{M}=\langle\mathcal{S},\{\mathcal{A}_n\},T,\{\mathcal{O}_n\},O,R,\gamma\rangle$. The details are provided below.

\subsection{Environment State Space (\texorpdfstring{$\mathcal{S}$}{S})} 
We define the environment state to include all variables required for a Markovian state transition. 
The environment state at time slot \(t\) is
\begin{equation}
\bm{s}(t) = \big( \bm{X}(t), \bm{q}(t), \bm{\Delta}(t) \big),
\label{eq:state_def}
\end{equation}
where $\bm{X}(t) = \big( X_{1}(t), X_{2}(t), \dots, X_{K}(t) \big)$ is the vector of fresh server availability states with $X_{k}(t) \in \{0,1\}$ defined in~\eqref{eq:server_status}, $\bm{q}(t) = \big( q_{1}(t), q_{2}(t), \dots, q_{K}(t) \big)$ is the vector of fresh queue lengths with $q_{k}(t) \in \{0,1,\dots,Q_k\}$ updated according to~\eqref{eq:queue_update}, and $\bm{\Delta}(t) = \big( \Delta_{n,k}(t) \big)_{n=1,k=1}^{N,K}$ is the AoI matrix for all $N$ dispatchers and $K$ servers evolving as in~\eqref{eq:aoi_update}.

\subsection{Action Space (\texorpdfstring{$\mathcal{A}$}{A})}
At each time slot \(t\), each dispatcher \(n\) selects a two-dimensional action 
\begin{equation}
a_{n}(t) = \big( \bm{c}_{n}(t), \bm{u}_{n}(t) \big),
\label{eq:per_dispatcher_action}
\end{equation}
where \(\bm{c}_{n}(t) = \big( c_{n,1}(t), \dots, c_{n,K}(t) \big)\) contains the binary server query decisions defined in \eqref{eq:query_def}, and \(\bm{u}_{n}(t) = \big( u_{n,1}(t), \dots, u_{n,K}(t) \big)\) contains the job-dispatch decisions, where \(u_{n,k}(t) = 1\) if dispatcher \(n\) assigns a newly arrived job to server \(k\) at time \(t\), and \(u_{n,k}(t) = 0\) otherwise. 
The joint action of all dispatchers at time \(t\) is 
\begin{equation}
\bm{a}(t) = \big( a_{1}(t), a_{2}(t), \dots, a_{N}(t) \big).
\label{eq:joint_action}
\end{equation}
Each dispatcher may query multiple servers in a single time slot but can dispatch at most one job, enforced by the constraint
\begin{equation}
\sum_{k=1}^{K} u_{n,k}(t) \leq 1, \quad \forall n \in \{1, \dots, N\}.
\label{eq:action_constraint}
\end{equation}

\subsection{State Transition Function (\texorpdfstring{$T$}{T})}
The environment state $\bm{s}(t)$ evolves according to the dynamics of the server availability, queue lengths, and AoI values. 
The fresh server states $\bm{X}(t)$ evolve as independent binary Markov chains with transition probabilities given by~\eqref{eq:transition_matrix}.  
The fresh queue lengths $\bm{q}(t)$ update according to~\eqref{eq:queue_update}, which depends on job arrivals, dispatching actions $\bm{u}(t)$, and server availability $\bm{X}(t)$.  
The AoI values $\bm{\Delta}(t)$ evolve according to~\eqref{eq:aoi_update}, driven by query actions $\bm{c}(t)$ and acknowledgment feedback $\bm{\alpha}(t)$.

\subsection{Observation Space (\texorpdfstring{$\mathcal{O}$}{O})}
At time slot $t$, dispatcher $n$ receives a local observation
\begin{IEEEeqnarray}{rCl}
\!\!\!\!\!\!o_{n}(t) &=& \nonumber \\
& & \!\!\!\!\!\!\!\!\!\!\!\!\! (X_k(t-\Delta_{n,k}(t)),\, q_k(t-\Delta_{n,k}(t)),\, \Delta_{n,k}(t))_{k=1}^K ,
\label{eq:observation_per_dispatcher}
\end{IEEEeqnarray}
based on the most recent feedback received from servers.  
The global joint observation state of all dispatchers is
\begin{equation}
\bm{o}(t) = \big( o_{1}(t), o_{2}(t), \dots, o_{N}(t) \big).
\label{eq:joint_observation}
\end{equation}

\subsection{Observation Function (\texorpdfstring{$O$}{O})}
The centralized critic uses the full state \(\bm{s}(t)\) in~\eqref{eq:state_def}, whereas each actor \(n\) receives only \(o_n(t)\) from~\eqref{eq:observation_per_dispatcher}. We model observations as a deterministic mapping,
\begin{IEEEeqnarray}{rCl}
\bm{o}(t{+}1) &=& h\!\big(\bm{s}(t{+}1),\, \bm{o}(t),\, \bm{a}(t)\big),
\end{IEEEeqnarray}
which induces the degenerate kernel
\begin{IEEEeqnarray}{rCl}
O(\bm{o}\mid \bm{s}', \bm{o}', \bm{a}) 
&=& \mathbf{1}\!\big\{\bm{o}=h(\bm{s}',\,\bm{o}',\,\bm{a})\big\}. 
\label{eq:obs_prob_func}
\end{IEEEeqnarray}

\subsection{Reward Function (\texorpdfstring{$R$}{R})}\label{sec:reward}
The objective is to maximize long-term job execution throughput while limiting the overhead of server queries. For dispatcher \(n\) at time \(t\), this trade-off is quantified by the instantaneous reward  
\begin{IEEEeqnarray}{rCl}\label{eq_rnt}
r_n(t) &=& \sum_{k=1}^K \sigma_{n,k}(t)
\;-\;
\beta \sum_{k=1}^K c_{n,k}(t),
\end{IEEEeqnarray}
where $\sigma_{n,k}(t) \in \{0,1\}$ is the job completion flag defined in~\eqref{eq:ack_def}, $c_{n,k}(t) \in \{0,1\}$ is the query decision in~\eqref{eq:query_def}, and $\beta \geq 0$ is the per-query cost.  
The system-level reward is defined as  
\begin{equation}
R(\bm{s}(t), \bm{a}(t)) = \sum_{n=1}^{N} r_n(t),
\label{eq:global_reward}
\end{equation}
with $\bm{s}(t) \in \mathcal{S}$ and $\bm{a}(t) \in \mathcal{A}$ denoting the environment state and joint action.  
The cooperative MA-POMDP problem for joint job dispatching and server querying across all $N$ dispatchers is formulated as   
\begin{equation}
\sup_{\{\bm a(t),\, t \geq 0\}} 
\mathbb{E} \left[ \sum_{t=0}^\infty \gamma^t R\big(\bm{s}(t), \bm{a}(t)\big) \right],
\label{eq:ma_objective}
\end{equation}
where \(\gamma \in [0,1)\) is the discount factor.

\section{Proposed MAPPO-Based Solution}
We adopt the Multi-Agent Proximal Policy Optimization (MAPPO) framework to solve the cooperative MA-POMDP formulated in Section~\ref{sec:pomdp}. The aim is to learn a decentralized policy $\pi^* = (\pi_1, \dots, \pi_N)$ that maximizes the team discounted return in~\eqref{eq:ma_objective}.

\subsection{PPO}
Proximal Policy Optimization (PPO)~\cite{schulman2017proximal} is a policy gradient algorithm that introduces a clipped surrogate objective to enable stable mini-batch updates and mitigate sensitivity to step size. Let $\pi_{\theta}$ and $V_{\phi}$ denote the actor and critic networks with parameters $\theta$ and $\phi$, respectively.
The clipped surrogate objective function of PPO is formulated as  
\begin{IEEEeqnarray}{rCl}
L_{\mathrm{CLIP}}(\theta) 
& = & \mathbb{E}_t \Big[ \min\big( \rho_t(\theta) \hat{A}_t,\; \nonumber\\
& & \quad\;\; \text{clip}(\rho_t(\theta), 1-\epsilon, 1+\epsilon) \hat{A}_t \big) \Big],
\label{eq:clip_loss}
\end{IEEEeqnarray}
where  
\begin{equation}
\rho_t(\theta) = \frac{\pi_{\theta}(a_t \mid o_t)}{\pi_{\theta_{\text{old}}}(a_t \mid o_t)}
\end{equation}
is the ratio of the new policy to the old policy, and $\epsilon > 0$ is the clipping parameter that limits excessive policy updates.  
The state–action value is given by  
\begin{equation}
Q(\bm{s}(t),\bm{a}(t)) = R(\bm{s}(t),\bm{a}(t)) + \gamma V_{\phi}(\bm{s}(t+1)),
\label{eq:Q_function}
\end{equation}
where $R(\bm{s}(t),\bm{a}(t)) = \sum_{n=1}^N r_n(t)$ is the team reward.
The term $\hat{A}_t$ is the empirical advantage estimate, computed using the Generalized Advantage Estimator (GAE)~\cite{schulman2015high} as  
\begin{equation}
\hat{A}_t \approx Q(\bm{s}(t),\bm{a}(t)) - V_{\phi}(\bm{s}(t)).
\label{eq:advantage_equation}
\end{equation}
The critic network is parameterized by $\phi$ and is learned using gradient descent with the loss function  
\begin{equation}
    L_{\text{Value}}(\phi) = \mathbb{E}_t\big[(V_\phi(\bm{s}(t)) - \hat{V}(t))^2\big],
\label{eq:value_loss}
\end{equation}
where $\hat{V}(t)$ is the \textit{Monte Carlo estimate} of the return starting from time step \(t\), defined as  
\begin{equation}
    \hat{V}(t) = \sum_{l=0}^{T - t} \gamma^l R(\bm{s}(t + l), \bm{a}(t + l)),
\label{eq:monte_carlo_return}
\end{equation}
with $\gamma \in [0,1)$ denoting the discount factor. Here, the dummy index \(l\) represents the number of steps ahead of the current time \(t\), and the sum accumulates discounted team rewards until the end of the episode at time \(T\). 
For the joint action \((\bm{c}_n, \bm{u}_n)\), the entropy is computed as \(\mathcal{H}(\pi_\theta^c)+\mathcal{H}(\pi_\theta^{u})\), assuming independent heads. The entropy loss is defined as:
\begin{equation}
    L_{\text{entropy}}(\theta) = - \mathbb{E}_t\big[\sum_{a\in\mathcal{A}} \pi_\theta^c(a \mid o(t)) \log \pi_\theta^c(a \mid o(t))\big].
\label{eq:entropy_loss}
\end{equation}
The total loss combines the policy term with value and entropy regularization terms to encourage exploration in early training and more deterministic behavior later. The total loss is defined as~\cite{schulman2017proximal}:
\begin{equation}
   L_{\text{total}}(\theta,\phi) = -L_{\text{CLIP}}(\theta) + c_v L_{\text{Value}}(\phi) - c_e L_{\text{entropy}}(\theta), 
\label{eq:total_loss}
\end{equation}
where $c_v > 0$ and $c_e > 0$ are weighting coefficients.

\subsection{MAPPO}
MAPPO extends PPO to cooperative multi-agent settings under the CTDE paradigm.

In training, a centralized critic \(V_{\phi}(\bm{s}(t))\) observes the full 
joint state \(\bm{s}(t)\) from~\eqref{eq:state_def}, while each actor 
\(\pi_{\theta_n}(a_n(t)\mid o_n(t))\) relies only on its local observation 
\(o_n(t)\) in~\eqref{eq:observation_per_dispatcher}. Each actor outputs a joint 
distribution over query and dispatch actions $(\bm{c}_n,\bm{u}_n)$, while the 
critic estimates the expected return. Both are trained with the global reward 
in~\eqref{eq:global_reward}. The advantage $\hat{A}_n(t)$ is computed as in 
\eqref{eq:advantage_equation}, with $Q_n$ from~\eqref{eq:Q_function}. Actor 
updates use the PPO clipped surrogate loss~\eqref{eq:clip_loss}, the critic 
uses the value loss~\eqref{eq:value_loss}, and the total loss 
\eqref{eq:total_loss} combines policy, value, and entropy terms.


During execution, the critic is unavailable. Each dispatcher $n$ applies its 
trained policy $\pi_{\theta_n}(a_n(t)\mid o_n(t))$ based only on its local 
observation. At each time slot, the dispatcher selects query actions 
$\bm{c}_n(t)$ and, if a job arrives, dispatches $\bm{u}_n(t)$. Training under 
CTDE ensures policies capture the influence of other agents and system dynamics, 
enabling effective decentralized decision-making.

The overall procedure follows the standard MAPPO framework adapted 
from~\cite{kang2023cooperative}, combining centralized critic updates with 
decentralized actor training. Our implementation uses the canonical MAPPO loop 
with clipped surrogate loss, centralized value updates, and entropy regularization.

\section{Numerical Results}

\begin{figure}[t]
    \centering
    \includegraphics[width=0.8\linewidth]{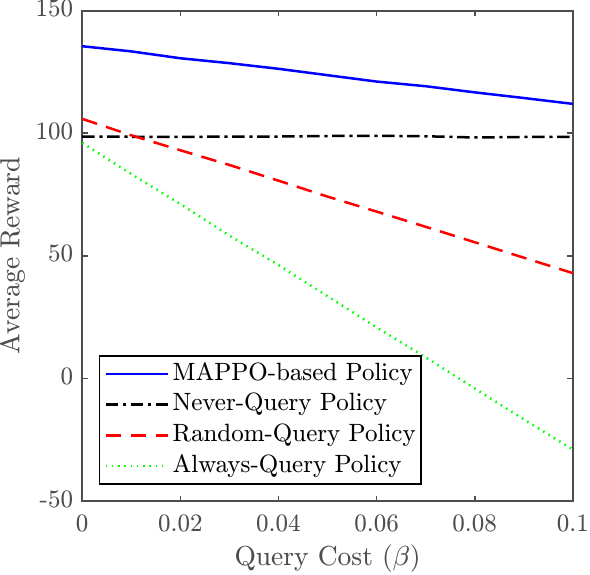}
    \caption{Average reward for varying query cost ($\beta$).}
    \label{fig:beta_sweep}
\end{figure}

\begin{figure}[t]
    \centering
    \includegraphics[width=0.8\linewidth]{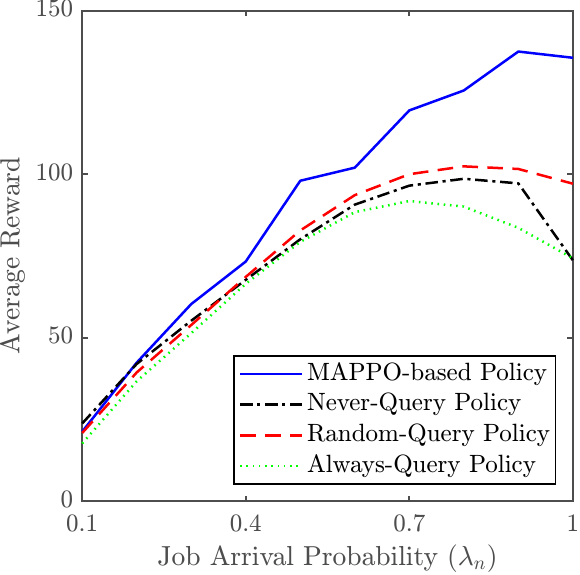}
    \caption{Average reward for varying job arrival probability ($\lambda_n$) for all dispatchers.}
    \label{fig:lambda_sweep}
\end{figure}

\begin{figure}[t]
    \centering
    \includegraphics[width=0.8\linewidth]{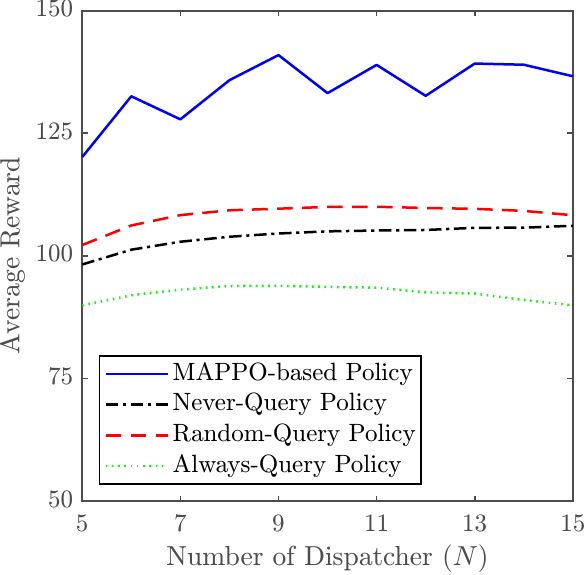}
    \caption{Average reward with increasing number of dispatchers ($N$).}
    \label{fig:n_sweep}
\end{figure}

To evaluate the effectiveness of our approach, we compare the MAPPO-based Policy against three baselines: (i) Never-Query Policy, which does not issue explicit queries and relies solely on execution feedback; (ii) Random-Query Policy, which queries each server independently with probability 0.5; and (iii) Always-Query Policy, which queries all servers in every slot. All policies adopt the same dispatch rule, assigning jobs to the least-loaded server given the information available. Unless otherwise specified, the default parameters are $K=5$ servers, $N=5$ dispatchers, job arrival probability $\lambda_n=0.8$, query cost $\beta=0.005$, and queue capacity $Q_k=3$. Server reliabilities are heterogeneous, with $(\phi_k,\psi_k)=(0.95,0.50)$ for odd-indexed servers and $(0.50,0.95)$ for even-indexed servers.

We first vary the query cost $\beta$ from 0 to 0.1. Figure~\ref{fig:beta_sweep} shows that the MAPPO-based Policy achieves up to 37\% higher rewards than Never-Query and over 160\% higher than Random across the entire cost spectrum. While MAPPO's reward fluctuates slightly with $\beta$, it remains robust even under high query costs by adapting query frequency. In contrast, the Always-Query Policy degrades rapidly with increasing cost and becomes negative at high costs, the Never-Query Policy remains flat, and the Random-Query Policy gradually declines. These results confirm MAPPO's ability to balance information freshness with communication cost.

Next, we vary the job arrival probability $\lambda_n$ from 0.1 to 1.0 for all dispatchers. As shown in Figure~\ref{fig:lambda_sweep}, at very low arrival rates ($\lambda = 0.1$) the Never-Query Policy slightly outperforms MAPPO, since stale information is sufficient. However, for $\lambda \geq 0.2$, MAPPO surpasses all baselines, with gains reaching up to 84\% over Never-Query and 40\% over Random under heavy traffic ($\lambda = 1.0$). This demonstrates MAPPO's ability to adaptively increase query frequency when timely information is critical.

Finally, we vary the number of dispatchers from $N=5$ to $N=15$. Figure~\ref{fig:n_sweep} shows that the MAPPO-based Policy scales effectively with system size. At larger $N$, MAPPO achieves up to 60\% higher rewards than the Always-Query Policy and approximately 40\% higher rewards than the Never-Query Policy, highlighting its effectiveness in coordinating distributed querying and dispatching decisions at scale.

\section{Conclusion}
In this paper, we addressed a goal-oriented communication problem for edge computing systems, where multiple dispatchers cooperatively assign jobs to shared servers with finite queues and time-varying availability. We formulated the task as a cooperative multi-agent decision-making problem and proposed a MAPPO-based policy that jointly optimizes query scheduling and job dispatching under partial and stale observations. Numerical evaluations demonstrate that the proposed cooperative MAPPO-based approach achieves superior throughput-cost tradeoffs and significantly outperforms conventional baselines across varying query costs, job arrival rates, and dispatchers.

\bibliographystyle{IEEEtran}
\bibliography{references}

\end{document}